\documentstyle[editedbook,epsfig,psfig,epsf]{mq}

\def\cm2{cm$^2$ }
\def\se1{s$^{-1}$ }

\newcommand{\gtsim}{\mbox
{{\raisebox{-0.4ex}{$\stackrel{>}{{\scriptstyle\sim}}$}}}}

\begin{opening}
\title{The ruff of equatorial emission around the SS433 jets: its
spectral index and origin}

\author{Katherine M.\ Blundell$^1$, Michael P.\ Rupen$^2$,
Amy J.\ Mioduszewski$^2$, \ \ }
\author{Tom W.\ B.\ Muxlow $^3$ \& Philipp
Podsiadlowski$^1$} 
\institute{$^1$ Oxford University Astrophysics, Keble Road, Oxford,
OX1 3RH, UK,\\ 
$^2$ NRAO, Socorro, New Mexico 87801, USA,\\
$^3$ Jodrell Bank Observatory, Cheshire SK11 9DL, UK}

\end{opening}

\runningtitle{Spectral index of equatorial emission around SS433}
\runningauthor{Blundell, Rupen, Mioduszewski, Muxlow \& Podsiadlowski}

\begin{document}
\vspace{-0.5cm}
\begin{abstract}
{\small We present unique radio observations of SS433, using MERLIN,
the VLBA, and the VLA, which allow us to, for the first time, properly
image and derive a meaningful spectral index for the `ruff' of
equatorial emission which surrounds SS433's jet.  We interpret
this smooth ruff as a wind-like outflow from the binary.}
\end{abstract}

\section{SS433's ruff of equatorial emission}

\begin{figure}[h]
\centering
\hbox{
\psfig{file=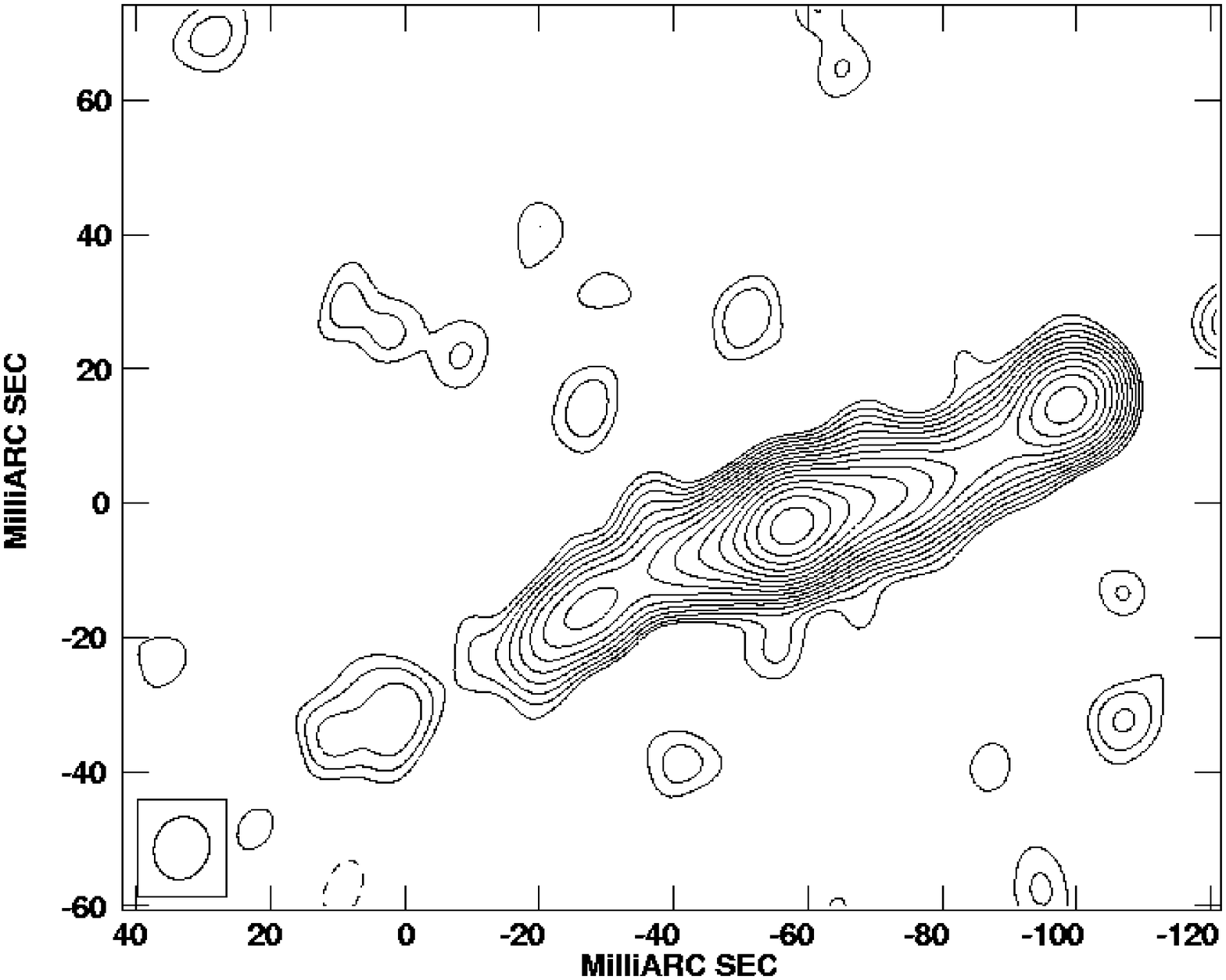,width=0.5\textwidth}
\hfill
\psfig{file=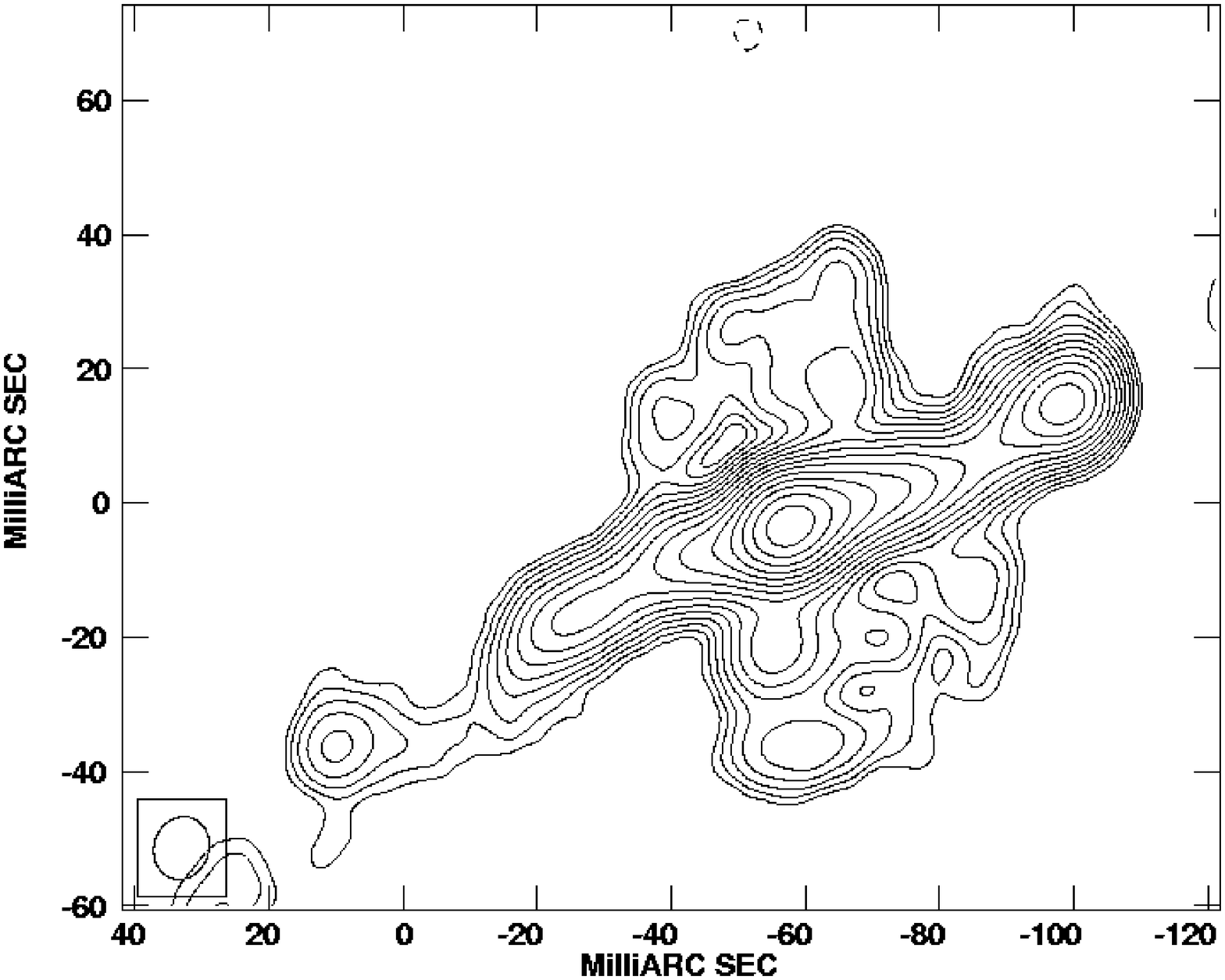,width=0.5\textwidth}
}
\caption{Two identically-contoured images of the central
quarter-arcsec of SS433 at 5\,GHz which have been made (left) {\bf
without} and (right) {\bf with} shorter baselines from MERLIN than the
VLBA's long baselines.  The right image clearly reveals a smooth ruff
of emission around SS433's jet which is only barely hinted at in the
left image.  The measured flux density of the ruff is low by an order
of magnitude if short baselines are missing, as quantified below:
\label{fig:ruff}}
\end{figure}

The central quarter-arcsecond of SS433's appearance at 5\,GHz is rich
in structure: both compact and smooth features may be found.  
To image this at radio wavelengths requires an interferometer
with sufficiently long baselines to give adequate resolution.  Those long
baselines will act as a spatial frequency filter which only detects
compact emission; they are insensitive to larger-scale structures.  At a
frequency such as 5\,GHz, short baselines are also needed to
faithfully detect smoother extended emission.  We illustrate this
in Figure~\ref{fig:ruff}: the left figure shows the central region of
SS433 imaged using only the VLBA; the right
figure shows the same region, with the same contour levels, on the
same epoch, at the same frequency, at the same resolution, using the
same VLBA data, but adding in also the shorter baselines of
MERLIN.  In the left figure the only believable brightness structure
is that associated with SS433's familiar jet, although hints of surrounding
emission are also seen.  On the right figure, a wide smooth structure
surrounding the jet appears, which we \cite{Blu01} have
termed SS433's {\em ruff}.  Since the spatial filtering depends on the
baseline length as measured in {\em wavelengths}, it is most severe at the
highest frequencies, and even the VLBA alone can detect the ruff at 1.4\,GHz
\cite{Blu01}.  

The spectral index of any extended emission may only be measured if
that emission has been properly sampled at both frequencies.
With VLBA data alone one can simply not detect the ruff at 5\,GHz,
while at 1.4\,GHz it is obvious; the undersampling at high frequencies
would lead to the derivation of a spuriously {\em steep}
spectral index.   Time variability is a further complication, making it
essential to observe the two frequencies simultaneously.
The observations we presented in
\cite{Blu01} at 1.4\,GHz and at 5\,GHz were taken on the same day 
(1998\,Mar\,7), and included the VLBA, MERLIN, and the VLA. 
This is thus a {\em unique} dataset: there are sufficient short baselines at
high frequency to adequately sample the emission, and both frequencies were
observed quasi-simultaneously.  We find a flat spectral index for the
anomalous emission (see below).  Paragi et al. \cite{Par02a} claim a steep
spectral index; but their
high-frequency data are undersampled (as they pointed out in \cite{Par99}),
and their observations at the different frequencies are not simultaneous.
Those data do not therefore usefully constrain the spectral index.

\begin{figure}[h]
\centering \hbox{ \psfig{file=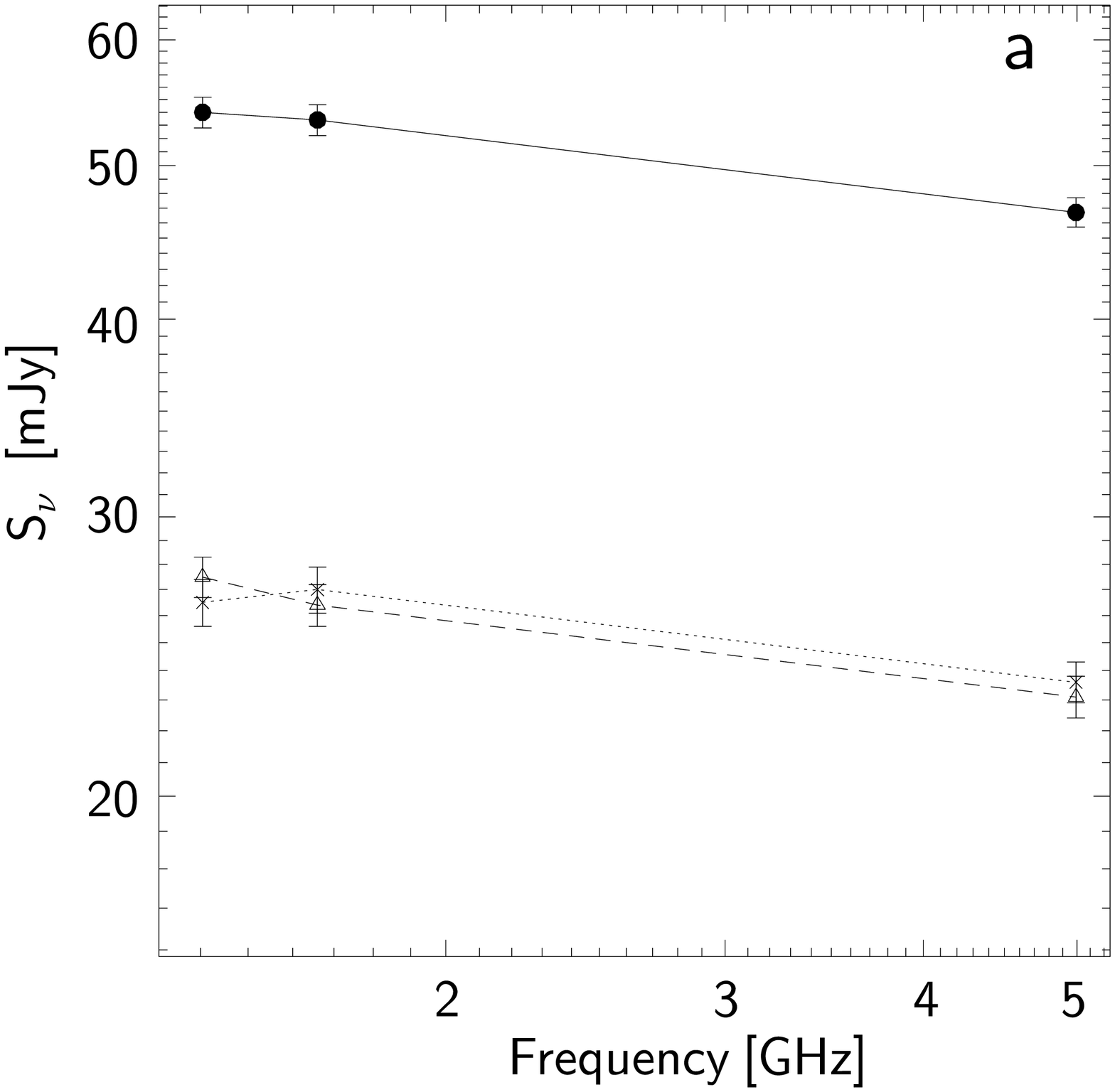,width=0.5\textwidth} \hfill
\psfig{file=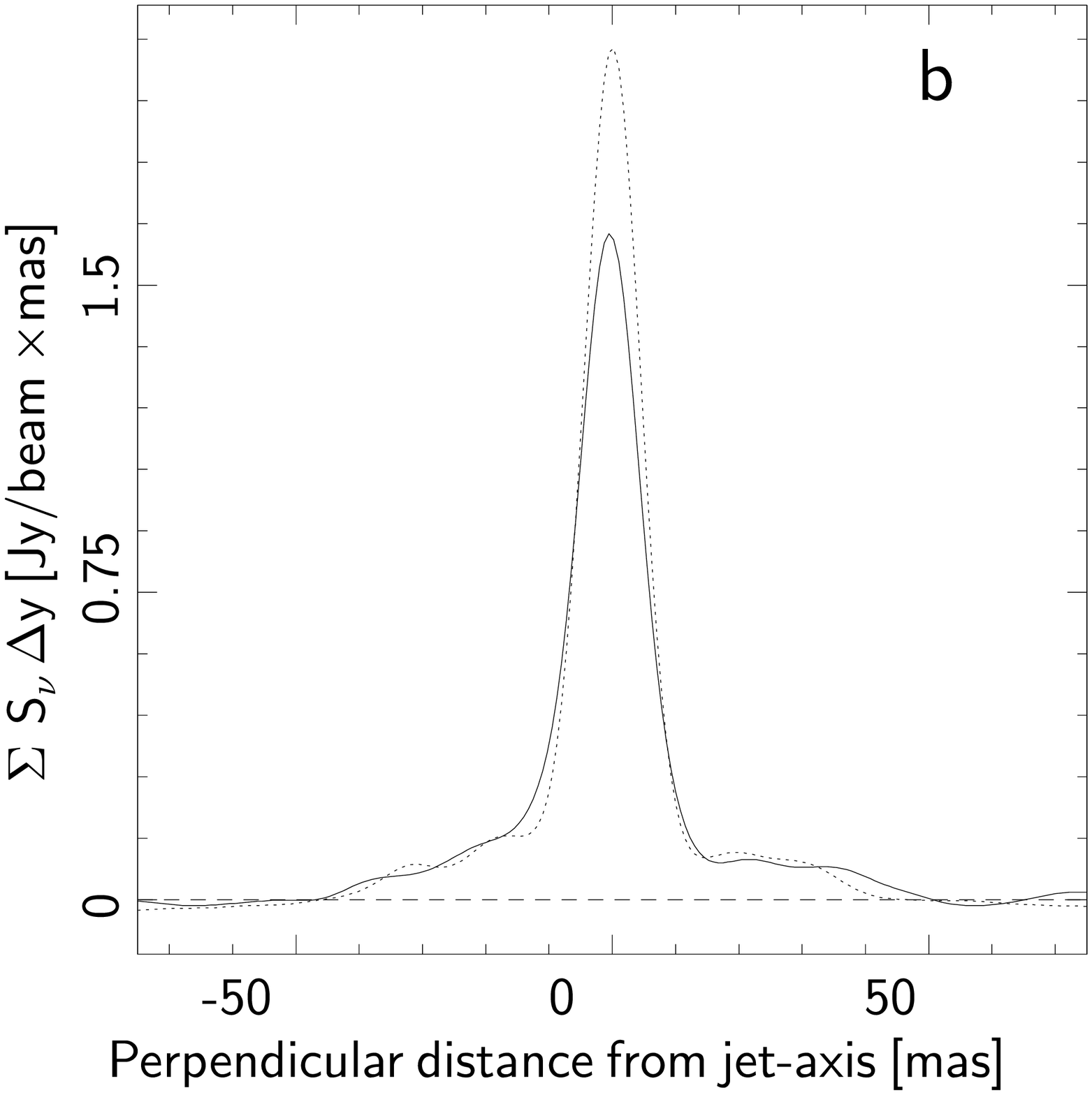,width=0.5\textwidth} }
\caption{{\em a:} The total flux density in the ruff emission as a
function of frequency (see text).  Crosses: northern emission;
triangles: southern emission; filled circles: sum of northern and
southern emission.  {\em b:} The flux density integrated over 40\,mas
strips parallel to the jet, as a function of distance perpendicular to
the jet, at 10\,mas resolution.  The solid line is 18\,cm, the dotted
line the 6\,cm data.  Note the flat spectrum of the ruff emission,
compared to the core.
\label{fig:alpha}}
\end{figure}

Our measurements of the distribution of the spectral index across SS433's
ruff are shown in Figure\,\ref{fig:alpha}.  The spectral indices of the ruff 
were measured after convolving our images to a common beam of $10
\times 10$\,mas$^2$ HPBW.  The resulting total flux densities,
measured in identical boxes in these images, which were chosen to
avoid the jet but include the full `ruff' emission, are shown in
Figure\,\ref{fig:alpha}$a$.  The spectral index for the combined
(northern+southern) emission is $\alpha=-0.12\pm0.02$ ($S_{\nu}
\propto \nu^\alpha$, where $S_{\nu}$ is the flux density at frequency
$\nu$). Most resolved synchrotron sources are characterized by $\alpha
< -0.4$; indeed, $\alpha=-0.1$ is normally considered the signature of
thermal bremsstrahlung emission as is often observed in outflows from
symbiotic binaries \cite{Sea84,Mik01}.  The complication here is that
the peak surface brightness corresponds to a brightness temperature of
$(2-4)\times10^7\,\rm K$ at 1.4\,GHz, implying a similar {\it lower
limit} to the physical temperature of a thermally-emitting plasma.

\begin{center}
\begin{tabular}{cll}
\hline
 & & \\[-0.1cm]
 & Measurements from  & Measurements from  \\
           & {\bf VLBA} only    &   {\bf VLBA \& MERLIN}  \\[0.2cm]
\hline
 & & \\[-0.1cm]
Northern ruff & \multicolumn{1}{c}{3.5 mJy} 
              & \multicolumn{1}{c}{22.8 mJy}    \\
Southern ruff & \multicolumn{1}{c}{3.1 mJy} 
              & \multicolumn{1}{c}{27.4 mJy}    \\[0.2cm]
              & \underline{\em without} short baselines 
              & \underline{\em with} short baselines \\[0.2cm]
\hline
\end{tabular}
\end{center}

The distribution of the flux density perpendicular to the jet is shown
in Figure\,\ref{fig:alpha}$b$, which suggests that the spectral index
is indeed almost flat throughout the ruff, and shows that the emission
extends to \gtsim\ $40\,\rm mas$ at our sensitivity, or
$\sim120\left(d/3\,\rm kpc\right)\,\rm AU$.  Note also that the ruff is
roughly symmetric about the jet.

\section{The origin of the smooth emission}

\def\Ms{\hbox{$\,M_{\odot}$}}
\def\yr{\hbox{$\,\hbox{yr}$}}

The most straightforward interpretation of the radio emission is that
it arises from mass outflow from the binary system that is enhanced
towards the orbital plane. Such mass loss could either (i) come from
the companion (most likely an O or B star), (ii) be a disc wind from
the outer parts of the accretion disc or (iii) arise from mass loss
from a proto-common envelope surrounding the binary components. The
detection of this mass loss will have important implications for our
understanding of the evolutionary state of this unique system. It has
been a long-standing puzzle how SS433 can survive so long in a phase
of extreme mass transfer ($\dot{M} \gtsim\ 10^{-5}\Ms\yr^{-1}$)
without entering into a common envelope phase where the compact object
spirals completely into the massive companion (for a recent discussion
see \cite{Kin00}).  Since the theoretically predicted mass-transfer
rate exceeds even the estimated mass-loss rate in the jets
($\dot{M}\sim 10^{-6}\Ms\yr^{-1}$; \cite{Beg80}), \cite{Kin00}
proposed that most of this transferred mass is lost from the system in
a radiation-pressure driven wind from the outer parts of the accretion
disc \cite{King99}.  A related problem exists in some
intermediate-mass X-ray binaries (IMXBs).  Models of the IMXB Cyg X-2
\cite{King99a,Pod00,Kolb00,Taur00} show that the system must have
passed through a phase where the mass-transfer rate was $\sim
10^{-5}\Ms\yr^{-1}$, exceeding the Eddington luminosity of the
accreting star by many orders of magnitude, without entering into a
common-envelope phase, and where almost all the transferred mass must
have been lost from the system. The observed emission in SS433
presented here may provide direct evidence of how such mass loss takes
place.


The existence of a disc-like outflow was first postulated by
\cite{Zwi91} to explain the variation with precession phase of the
secondary minimum in the photometric light curve.  \cite{Fab93}
proposed a disc-like expanding envelope caused by mass-loss from the
outer Lagrangian point L2 to explain the blue-shifted absorption lines
of H\,I, He\,I and Fe\,II (see also \cite{Mam80}, whose spectrum
shows that all the emission lines seen in SS433 have P-Cygni
profiles indicating the presence of outflowing gas).  \cite{Fil88}
observe remarkable double peaked Paschen lines,
with speeds close to 300\,${\rm km\,s^{-1}}$.

A rough estimate for the
mass-outflow rate, described in our paper \cite{Blu01} is:
\begin{eqnarray} 
\dot{M}&\simeq& 1.6\times 10^{-4}\,M_{\odot}\,\mbox{yr}^{-1}\,\,
\\
&&\hspace{1cm}\times S_{50}^{3/4}\,d_{3}^{3/2}\,v_{300}\,\nu_{1.4}^{-1/2}\,
\bar{g}_{10}^{-1/2}\, (\sin\alpha)_{30}^{1/4},\nonumber 
\end{eqnarray}  
where $S_{50} = S_\nu/50$\,mJy, $d_{3}=d/3\,$kpc, $v_{300}=
v_\infty/300\,$km\,s$^{-1}$, $\nu_{1.4}=\nu/1.4\,$GHz, $\bar{g}_{10}=
\bar{g}/10$ ($\bar{g}$ is the Gaunt factor for free-free emission,
$(\sin\alpha)_{30}=\sin\alpha/\sin 30^\circ$).

One of the major uncertainties in this estimate is the velocity of the
outflow, though a velocity of $\sim 300\,$km\,s$^{-1}$ is similar to
that of the lines seen by \cite{Fil88} and is close to the characteristic
orbital velocity of SS433, as one might expect for an outflow from the
binary system rather than either binary component.  Furthermore, if
this outflow started soon after the supernova explosion which formed
the compact object $\sim 10^4\,$yr ago and whose impressively circular
remnant is seen clearly in the images of \cite{Dub98}, a velocity of
$\sim 300\,$km\,s$^{-1}$ implies an extent of the outflow of $\sim
3\,$arcmin (for $d=3\,$kpc). Indeed, this is exactly the size of the
extended smooth emission seen by \cite{Dub98} and suggests that this
may be the outer extent of the same outflow.  The inferred mass-loss
rate, $\dot{M}\sim 10^{-4}\,M_{\odot}\,$yr$^{-1}$, is much higher than
any reasonable mass-loss rate from an O-star primary and suggests that
it is connected with the unusual short-lived phase SS433 is
experiencing. It could be mass loss from a common envelope that has
already started to form around the binary, or a hot coronal wind from
the outer parts of the accretion disc driven, e.g., by the X-ray
irradiation from the central compact source.

\section*{Acknowledgments}
K.M.B.\ thanks the Royal Society for a University Research Fellowship.
We warmly thank the conference organisers for a very stimulating
meeting.


\begin{thebibliography}{}

\bibitem[Begelman et al(1980)]{Beg80}
Begelman, M.C., Hatchett, S.P., McKee, C.F., Sarazin, C.L., Arons, J.,
1980, ApJ, 238, 722

\bibitem[Blundell et al(2001)]{Blu01}
Blundell, K.M., Mioduszewski, A.J., Muxlow, T.W.B.,
Podsiadlowski, Ph., Rupen, M.P., 2001, ApJ, 562, L79

\bibitem[Dubner et al.(1998)]{Dub98}
Dubner, G.M., Holdaway, M., Goss, W.M., \& Mirabel, I.F., 1998, AJ, 116,
1842  

\bibitem[Fabrika(1993)]{Fab93}
Fabrika, S.N., 1993, MNRAS, 261, 241

\bibitem[Filippenko et al.(1988)]{Fil88}
Filippenko, A.V., Romani, R.W., Sargent, W.L.W., \& Blandford,
R.D., 1988, AJ, 96, 242

\bibitem[King \& Begelman(1999)]{King99}
King, A.R.\ \& Begelman, M.C.\ 1999, ApJ, 519, L169

\bibitem[King \& Ritter(1999)]{King99a}
King, A.R.\ \& Ritter, H.\ 1999, MNRAS, 309, 253

\bibitem[King, Taam \& Begelman(2000)]{Kin00}
King, A.\ R., Taam, R.\ E., \& Begelman, M.\ C.\ 2000, ApJ, 530, {\sc l}25 

\bibitem[Kolb et al.(2000)]{Kolb00}
Kolb, U., Davies, M, King, A., \& Ritter, H.\ 2000, MNRAS, 317, 438

\bibitem[Mammano, Ciatti \& Vittone(1980)]{Mam80}
Mammano, A., Ciatti, F., \& Vittone, A.\ 1980, A\&A, 85, 14

\bibitem[Miko{\l}ajewska \& Ivison(2001)]{Mik01}
Miko{\l}ajewska, J.\ \& Ivison R.J., 2001, MNRAS, 324, 1023

\bibitem[Paragi et al(1999)]{Par99}
Paragi Z., Vermeulen R.C., Fejes I., Schilizzi R.T.,
Spencer R.E., \& Stirling A.M., 1999, A\&A, 348, 910

\bibitem[Paragi et al(2002a)]{Par02a}
Paragi Z., Fejes I., Vermeulen R.C., Schilizzi R.T.,
Spencer R.E., \& Stirling A.M., 2002a, in  Proc. 6th European VLBI
Network Symposium, eds Ros E., Porcas R.W., Lobanov A.P. and Zensus
J.A, astro-ph/0207061

\bibitem[Paragi et al(2002b)]{Par02b}
Paragi Z., Fejes I., Vermeulen R.C., Schilizzi R.T.,
Spencer R.E., \& Stirling A.M., 2002n, these proceedings;
astro-ph/0208125  

\bibitem[Podsiadlowski \& Rappaport(2000)]{Pod00}
Podsiadlowski, Ph.\ \& Rappaport, S. 2000, ApJ, 529, 946


\bibitem[Seaquist, Taylor, \& Button(1984)]{Sea84}
Seaquist, E.R., Taylor, A.R., \& Button, S. 1984, ApJ, 284, 202

\bibitem[Tauris, van den Heuvel, \& Savonije(2000)]{Taur00}
Tauris, T.M., van den Heuvel, E.P.J., \& Savonije, G.J.\ 2000, ApJ, 530,
L93

\bibitem[Zwitter, Calvani, \& D'Odorico(1991)]{Zwi91}
Zwitter, T., Calvani, M., \& D'Odorico, S.\ 1991, A\&A, 251 92



\end{thebibliography}
\end{document}